\documentclass[12pt]{article}
\usepackage{amsmath,amssymb,amsfonts,mathrsfs}
\usepackage{graphicx}

\usepackage{amsfonts,amsmath,xspace,amsmath,varioref,ifpdf} %

\IfFileExists{srcltx.sty}{\usepackage[active]{srcltx}}

\def\hybrid{\topmargin 0pt      \oddsidemargin 0pt
           \headheight 0pt \headsep 0pt
          \voffset=-0.5cm         
           \textwidth 6.25in       
           \textheight 9.5in       
           \marginparwidth 0.0in
           \parskip 5pt plus 1pt   \jot = 1.5ex}
\catcode`\@=11

\newcount\hour
\newcount\minute
\newtoks\amorpm
\hour=\time\divide\hour by60 \minute=\time{\multiply\hour by60
\global\advance\minute by-\hour}
\edef\standardtime{{\ifnum\hour<12 \global\amorpm={am}%
           \else\global\amorpm={pm}\advance\hour by-12 \fi
           \ifnum\hour=0 \hour=12 \fi
           \number\hour:\ifnum\minute<10 0\fi\number\minute\the\amorpm}}
\edef\militarytime{\number\hour:\ifnum\minute<10 0\fi\number\minute}

\def\draft{
  \oddsidemargin -.5truein \def\@oddfoot{{\scriptsize \it \underline{File
        \jobname.tex}. $Revision: 1.7 $} \hfil -- \footnotesize \thepage\ --
    \hfil{\scriptsize\it\today\quad\militarytime}}
  \let\@evenfoot\@oddfoot %
  \overfullrule 3pt %
  \let\label=\draftlabel %
  \let\marginnote=\draftmarginnote %
  \def\@eqnnum{(\theequation)%
    \rlap{\kern\marginparsep\scriptsize\bfseries\@eqnlabel}%
    \global\let\@eqnlabel\@vacuum}
}

\def\numberbysection{\@addtoreset{equation}{section}
           \def\theequation{\thesection.\arabic{equation}}}
\catcode`@=12 \relax
\hybrid \numberbysection

\begin{document}

\title{Time-dependent correlation function of the Jordan-Wigner operator as a Fredholm determinant}

\author{M. B.
Zvonarev\thanks{DPMC-MaNEP, University of Geneva, 24 quai Ernest-Ansermet, 1211 Geneva 4, Switzerland}, V. V. Cheianov\thanks{Department of Physics, Lancaster University, Lancaster,
LA1 4YB, United Kingdom}, T. Giamarchi\thanks{DPMC-MaNEP, University of Geneva, 24 quai Ernest-Ansermet, 1211 Geneva 4, Switzerland}}

\maketitle{}

\abstract{We calculate a correlation function of the Jordan-Wigner operator in a class of free-fermion models formulated on an infinite one-dimensional lattice. We represent this function in terms of the determinant of an integrable Fredholm operator, convenient for analytic and numerical investigations. By using Wick's theorem, we avoid the form-factor summation customarily used in literature for treating similar problems.}

\section{Introduction and main results}

\begin{figure}
\begin{center}
\includegraphics[width=8cm]{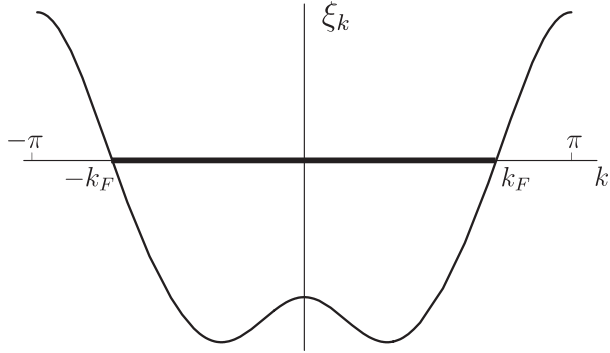}
\end{center}
\caption{Shown is an example of the single-particle energy $\xi_k$ as the function of quasimomentum $k$ in the first Brillouin zone, $-\pi\le k\le \pi$ (solid line). For the ground state average standing in Eq.~\eqref{corr} the $k$-space states between $-k_F$ and $k_F$ are occupied with fermions. This is illustrated with the bold line.}
\label{fig:fig1}
\end{figure}

In this paper we investigate the correlation function
\begin{equation}
D(\lambda;n,t)= \langle T \mathcal{O}_j(t) \mathcal{O}^\dagger_{j^\prime}(0) \rangle, \qquad n=j-j^\prime \label{corr}
\end{equation}
of the Jordan-Wigner operator
\begin{equation}
\mathcal{O}_j= e^{i\lambda\mathcal{N}_{j}}, \label{JW}
\end{equation}
where
\begin{equation}
\mathcal{N}_j = \sum_{m=-\infty}^j \varrho_m, \qquad \varrho_m= c^\dagger_m c_m. \label{N}
\end{equation}
The operator $c^\dagger_m$ ($c_m$) is the fermion creation (annihilation) operator at the $m$-th lattice cite of an infinite one-dimensional (1D) lattice. These operators obey canonical anticommutation relation, $\{c_m,c^\dagger_{m^\prime}\}=\delta_{mm^\prime}$. The symbol $T$ in Eq.~\eqref{corr} stands for the time ordering. The function~\eqref{corr} is a $2\pi$-periodic function of a real parameter $\lambda.$ The time evolution of the operators in Eq.~\eqref{corr} is generated by the free-fermion Hamiltonian
\begin{equation}
H= \int_{-\pi}^\pi \frac{dk}{2\pi} \xi_k \tilde c^\dagger_k \tilde c_k, \qquad \tilde c_k=\sum_m e^{-ikm}c_m, \label{ff}
\end{equation}
where $\xi_k$ is the single-particle energy as a function of the Bloch quasimomentum in the first Brillouin zone, $-\pi\le k\le \pi.$ For simplicity we assume that the equation $\xi_k=0$ has only two solutions $k=\pm k_F.$ In this case, the ground state average $\langle\cdots\rangle$ in Eq.~\eqref{corr} is taken over the Fermi sea of fermions filling the single particle states between $-k_F$ and $k_F:$
\begin{equation}
n_F(k)= \langle \tilde c^\dagger_k \tilde c_k \rangle= \left\{ \begin{array}{ll} 1, & k\in[-k_F,k_F]
\\ 0, &  \text{otherwise} \end{array} \right. . \label{nF}
\end{equation}
An example of $\xi_k$ satisfying Eq.~\eqref{nF} is shown in Fig.~\ref{fig:fig1}.

The goal of the paper is to derive a Fredholm determinant representation for the function~\eqref{corr}. This representation was used in our recent paper~\cite{zvonarev_BoseHubb08} to investigate correlation functions in the 1D Bose-Hubbard model with spin. However, in Ref.~\cite{zvonarev_BoseHubb08} we neither give the explicit expressions~\eqref{main0}--\eqref{zdef} nor explain a way they can be obtained. It is important to stress that the function~\eqref{corr} appears not only in the Bose-Hubbard model with spin. In the regime of strong interparticle repulsion it is possible for a broad class of models carrying spin and charge degrees of freedom to factorize spin and charge dynamics in such a way that the charge part is fully described by the function~\eqref{corr}. This was demonstrated in Refs.~\cite{ogata_inf, aristov_spinfluct_97, matveev_asymm_zero_bias07, matveev_heisenberg_electrons07, akhanjee_ferrobosons07, matveev_isospin_bosons08, zvonarev_BoseHubb08} by representing the original correlation function as a convolution of the charge correlation function~\eqref{corr} with some spin correlation function of the Heisenberg model (see, for example, Eq.~(3.5) of Ref.~\cite{ogata_inf}, Eq.~(8) of Ref.~\cite{aristov_spinfluct_97}, Eq.~(11) of Ref.~\cite{zvonarev_BoseHubb08}, etc.). The authors~\cite{ogata_inf, aristov_spinfluct_97, matveev_asymm_zero_bias07, matveev_heisenberg_electrons07, akhanjee_ferrobosons07, matveev_isospin_bosons08} do not give Fredholm determinant representation for the function~\eqref{corr}, analyzing it by other methods. An approach complementary to the one of Refs.~\cite{zvonarev_BoseHubb08, ogata_inf, aristov_spinfluct_97, matveev_asymm_zero_bias07, matveev_heisenberg_electrons07, akhanjee_ferrobosons07, matveev_isospin_bosons08} is used in Refs.~\cite{izergin_impenetrable_fermions, izergin_impenetrable_hubbard}. There 1D gas of bosons (and fermions) with spin is considered in a limit of infinitely strong short-range interparticle repulsion. The explicit expressions for the correlation functions are obtained by summing up matrix elements of the transitions to the intermediate states (form-factors summation technique). The results have the form of a Fredholm determinant convolved with some weight function. When compared with the predictions of Refs.~\cite{zvonarev_BoseHubb08, ogata_inf, aristov_spinfluct_97, matveev_asymm_zero_bias07, matveev_heisenberg_electrons07, akhanjee_ferrobosons07, matveev_isospin_bosons08} these results can be unfolded to the form of the convolution of some spin correlation function with the function~\eqref{corr}. This way one can indeed get the Fredholm determinant representation for the function~\eqref{corr}. However, in our opinion, it is easier to calculate the determinant directly, as explained in our paper. We note that correlation functions similar to the one defined by Eq.~\eqref{corr} appear in the analysis of many other important 1D systems. The number of works on the subject is large and we, unfortunately, cannot give the exhaustive list of references here. They include the studies of the Tonks-Girardeau gas~\cite{schultz_TG63,lenard_TG64, korepin_TG_Fredholm90, korepin_book}, impenetrable anyons~\cite{patu_LenardI08, patu_LenardII08}, XY Heisenberg chain~\cite{lieb_spinchains, McCoy_Ising83, Perk_Ising_Tfinite84, colomo_XX0_Fredholm92, colomo_XX0_Fredholm93, Its_XY93}, non-linear quantum hydrodynamics~\cite{imambekov_universal08}, and the full counting statistics of quantum transport~\cite{levitov_fcs96}.

The Fredholm determinant representation for the correlation function~\eqref{corr} reads
\begin{equation}
D(\lambda;n,t)= \det (\hat I+\hat K),
\label{main0}
\end{equation}
where $\hat K$ is a linear integral operator
\begin{equation}
(\hat K f)(k) = \int_{-k_F}^{k_F} \frac{dp }{2\pi} K(k,p) f(p) \label{Kdef0}
\end{equation}
with the integral kernel
\begin{equation}
K(k,p)=2 \frac{ \ell_{+}(k) \ell_-(p) -  \ell_{-}(k) \ell_+(p) }
{\tan \frac{k-p}{2}} - 2 i \left[\ell_+(k) \ell_-(p) - \ell_-(k) \ell_+(p) \right] \label{Kdef}
\end{equation}
defined on $[-k_F,k_F]\times[-k_F,k_F].$ The functions entering this kernel are defined as follows:
\begin{equation}
\ell_{-}(k)=\frac{1}{\sqrt 2} \frac{1}{z(k)}
\end{equation}
and
\begin{equation}
\ell_{+}(k)=\frac{1}{\sqrt 2} \left[\frac{\sin\lambda}{2}z(k) + \frac{1-\cos\lambda}{2}
\frac{E(k)}{z(k)}  \right],
\end{equation}
where
\begin{equation}
E(k) =\int_0^{2\pi} \frac{dq}{2\pi}  \frac{z^2(q)-z^2(k)}
{\tan\frac{q-k}{2}}
\label{Edef}
\end{equation}
and
\begin{equation}
z(k)=e^{\frac{i}{2} (n k-\xi_k t) }. \label{zdef}
\end{equation}
The linear integral operator with the kernel \eqref{Kdef} belongs to the class of integrable operators, introduced in Refs.~\cite{Its_BoseInt90, Its_DE90}, see also Ref.~\cite{korepin_book} for further discussion. For the particular case of $\lambda=\pi$ the kernel~\eqref{Kdef} appears in Refs.~\cite{colomo_XX0_Fredholm92, colomo_XX0_Fredholm93, Its_XY93} for the XY Heisenberg chain and in Ref.~\cite{korepin_TG_Fredholm90} for the impenetrable Bose gas. The integrable operator obtained in Refs.~\cite{izergin_impenetrable_fermions, izergin_impenetrable_hubbard} coincide with ours up to a finite rank operator.
The continuum limit of the kernel~\eqref{Kdef} is encountered in the theory of impenetrable anyons~\cite{patu_LenardII08}.

In some of the above mentioned works (for example, in
Ref.~\cite{lieb_spinchains}, see also Ref.~\cite{Perk_Ising_Tfinite84} for the
list of references and detailed discussion) Wick's theorem is used to get the
determinant representation of correlation functions, in others form-factor
summation technique is employed. Within the form-factor approach the calculations have to be carried out in a finite system, and the thermodynamic limit is applied to the end result only. In contrast, Wick's theorem approach makes possible to work in the thermodynamic limit from the very beginning. Though less rigorous, the latter approach enjoys the advantage of being minimalist. In particular, it avoids constructing the determinant representation for the function~\eqref{corr} in the system with a finite number of particles and, therefore, significantly simplifies the presentation as compared with the form-factor approach. This motivates the use of the Wick's theorem in the present work.

\section{Derivation of the results}

Consider a countable set $\mathcal M$ of points in space-time. For brevity we shall use the notation $\varrho(x) \equiv \varrho_j(t),$ $c(x) \equiv c_j(t)$ and $c^\dagger(x) \equiv c_j^\dagger (t)$  for the operators at the point $x=(j,t)$ from $\mathcal M.$ For an arbitrary function $\gamma: \mathcal{M}\to \mathbb{C}$ consider the average
\begin{equation}
\langle T \prod_{x\in {\mathcal M}}
[1+\gamma(x)\varrho(x)] \rangle=
1+\sum_{Q=1}^\infty \sum_{\mathcal M_Q\subset \mathcal M} \langle T \prod_{x\in {\mathcal{M}_Q}} \gamma(x) c^\dagger(x) c(x) \rangle, \label{TT1}
\end{equation}
where the inner sum is taken over all distinct subsets $\mathcal M_Q$ of $\mathcal M$ containing $Q$ elements. Applying Wick's theorem to the average on the right hand side of Eq.~\eqref{TT1} we get
\begin{equation}
\langle T \prod_{x\in {\mathcal M_Q}} \gamma(x) c^\dagger(x) c(x) \rangle=
\det M(\mathcal{M}_Q).
\label{useful0}
\end{equation}
In this expression $\det M(\mathcal{M}_Q)$ denotes the determinant of a $Q
\times Q$ matrix, which for a given enumeration $x_1, \dots, x_Q$ of the elements of $\mathcal M_Q$ is defined by
\begin{equation}
[M(\mathcal M_Q)]_{a,b}= \gamma(x_a) G(x_a-x_b), \quad a,b=1, \ldots, Q.
\end{equation}
Here
\begin{equation}
G(x)\equiv G(j,t)=- \langle T c_j(t) c_0^\dagger\rangle, \qquad
\qquad G(j, 0)\equiv \lim_{\epsilon\to 0} G(j, -\epsilon)
\label{G0}
\end{equation}
is the $T$-ordered single-particle Green's function.
Combining Eqs.~\eqref{G0} and \eqref{useful0} we find
\begin{equation}
\langle T \prod_{x\in {\mathcal M}}
[1+\gamma(x)\varrho(x)] \rangle=\det(\hat I+\hat K).
\label{identity1}
\end{equation}
where $\hat K$ is a linear endomorphism of the linear space
$\mathscr V(\mathcal M)$ of complex-valued functions on $\mathcal M:$
\begin{equation}
(\hat K f)(x)=\sum_{x'\in \mathcal M}\gamma(x) G(x-x') f(x'), \qquad  f\in \mathscr V(\mathcal M).
\end{equation}

Let us now apply the formula~\eqref{identity1} to the correlation function \eqref{corr}. Using the identity
$\varrho^2_j=\varrho_j$  following from the anti-commutativity of
fermionic fields we get
\begin{figure}
\begin{center}
\includegraphics[width=8cm]{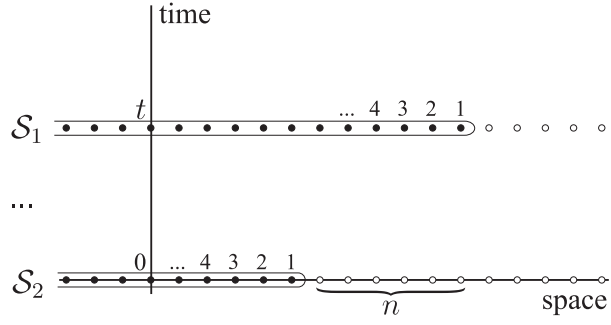}
\end{center}
\caption{Shown are two sets of points: $\mathcal S_1,$ Eq.~\eqref{string1}, and $\mathcal S_2,$ Eq.~\eqref{string2}, and the enumeration of their elements used in Eq.~\eqref{K}.}
\label{fig:fig2}
\end{figure}
\begin{equation}
e^{i\lambda \varrho_j }= 1+ \gamma \varrho_j, \qquad \gamma \equiv e^{i\lambda}-1.
\label{gammadef}
\end{equation}
The equation \eqref{corr} is then written as
\begin{equation}
D (\lambda; n,t)=\langle T \prod_{x\in \mathcal S_1}[1+ \gamma \varrho(x) ] \prod_{x\in \mathcal S_2} [1+ \gamma^* \varrho(x) ] \rangle,
\label{Dim}
\end{equation}
where
\begin{eqnarray}
& &\mathcal S_1= \left\{x: \; x=(n+1-a, t), \; a\in \mathbb N^* \right\}, \label{string1}\\
& &\mathcal S_2= \left\{x: \;  x=(1-a, 0), \;a\in \mathbb N^* \right\}
\label{string2}
\end{eqnarray}
and
\begin{equation}
\mathbb N^*= 1,2,3,\ldots .
\end{equation}

The right hand side of Eq.~\eqref{Dim} can be expressed as a determinant, Eq.~\eqref{identity1}, where $\hat K$ is a linear endomorphism
of $\mathscr V(\mathcal \mathcal S_1\cup \mathcal S_2)= \mathscr V(\mathcal \mathcal S_1) \oplus \mathscr V(\mathcal S_2).$ We represent $\hat K$ as a $2\times 2$ matrix
\begin{equation}
\hat K= \left(
\begin{array}{cc} \hat K^{11} & \hat K^{12} \\
\hat K^{21} & \hat K^{22}
\end{array} \right), \qquad K^{\alpha\beta}: \mathscr V(\mathcal S_\alpha) \to \mathscr V(\mathcal S_\beta), \qquad \alpha,\beta=1,2. \label{kmat1}
\end{equation}
Using the enumeration of elements of $\mathcal S_1$ and $\mathcal S_2$ defined in Eqs.~\eqref{string1} and \eqref{string2} and shown in Fig.~\ref{fig:fig2} we write a matrix element of the representation~\eqref{kmat1}:
\begin{equation}
[\hat K]_{a,b} =\left(\begin{array}{cc}
\gamma^* G(b-a)  & \gamma^* G(b-a-n,-t) \\
\gamma G(b-a+n, t) &  \gamma G(b-a)
\end{array}\right), \qquad a,b\in \mathbb N. \label{K}
\end{equation}
It is convenient to extend this definition to $a,b\in \mathbb Z$ by introducing
\begin{equation}
[\hat U]_{a,b} = \left\{ \begin{array}{ll} K_{a+1,b+1}, &
\quad  (a,b) \in \mathbb Z^+ \times \mathbb Z^+ \\ 0, &  \quad (a,b)\in ( \mathbb Z\times \mathbb Z ) {\backslash} ( \mathbb Z^+\times \mathbb Z^+) \end{array} \right. , \label{U}
\end{equation}
where
\begin{equation}
\mathbb Z^+ = 0,1,2,3,\ldots  \qquad \mathbb Z = 0,\pm1,\pm2,\pm3,\ldots .
\end{equation}
Since $\det (\hat I+\hat U)=\det (\hat I+\hat K)$ we get for Eq.~\eqref{Dim}
\begin{equation}
D(\lambda;n,t)=\det (\hat I+\hat U). \label{detIU}
\end{equation}

Defining the Fourier transform by
\begin{equation}
f(j)=\int_{-\pi}^{\pi} \frac{dk}{2\pi} e^{ikj} \tilde f(k) , \qquad \tilde f(k)=
\sum_{j=-\infty}^{\infty} e^{-ikj} f(j) \label{fourier}
\end{equation}
and applying it to Eq.~\eqref{U} we express $\hat U$ as a linear integral operator
\begin{equation}
(\hat U \tilde f)(k)= \int_{-\pi}^\pi \frac{dk}{2\pi} U(k,p) \tilde f(p)
\end{equation}
with the integral kernel
\begin{equation}
U(k,p)=\int_{-\pi}^{\pi} \frac{dq}{2\pi} \sum_{a, b=0}^\infty
e^{-iq(a-b)} e^{ika} e^{-ipb}  u(q), \label{Ukp}
\end{equation}
where
\begin{equation}
u (q)=\left(\begin{array}{cc}
\gamma^* \tilde G(q)  & \gamma^* \tilde G(q,-t)e^{-inq} \\
\gamma \tilde G(q, t)e^{i nq} &  \gamma \tilde G(q)
\end{array}\right). \label{uq}
\end{equation}
The function $\tilde G$ entering Eq.~\eqref{uq} is the Fourier transform of Green's function~\eqref{G0}:
\begin{equation}
\tilde G(q,t)=-\theta(t>0) [1-n_F(q)]e^{-i\xi_q t}+ \theta(t<0) n_F(q) e^{-i\xi_q t},
\end{equation}
where $\xi_q$ is the single-particle energy, Eq.~\eqref{ff},  and
$n_F(q)$ is the Fermi-Dirac function~\eqref{nF}. Performing the summation in Eq.~\eqref{Ukp} we obtain
\begin{equation}
U (k,p)= \int_{-\pi}^{\pi} \frac{dq}{2\pi} \eta(k-q)\eta(q-p) u (q),
\label{Mkernel}
\end{equation}
where
\begin{equation}
\eta(k)=\frac{1}{1-e^{ik-\epsilon}}
\end{equation}
and $\epsilon$ is an infinitesimal positive constant.

Next, we note that
\begin{equation}
u(q)= v(q)+  w(q), \label{uq2}
\end{equation}
where
\begin{equation}
 v(q)=
\left(\begin{array}{cc}
0  & 0  \\
- \gamma z^2(q) &  0
\end{array}\right) \label{vdef}
\end{equation}
and
\begin{equation}
w (q)= n_F(q)
\left(\begin{array}{cc}
\gamma^* & \gamma^* z^{-2}(q)  \\
\gamma z^2(q) &  \gamma
\end{array}\right).
\end{equation}
The function $z(q)$ is defined by Eq.~\eqref{zdef}. Using the decomposition \eqref{uq2} we write $\hat U=\hat V+ \hat W,$
where
\begin{equation}
V(k,p)= \int_{-\pi}^{\pi} \frac{dq}{2\pi} \eta(k-q)\eta(q-p)  v (q)
\end{equation}
and
\begin{equation}
W(k,p)= \int_{-\pi}^{\pi} \frac{dq}{2\pi} \eta(k-q)\eta(q-p) w (q).
\end{equation}
It follows from Eq.~\eqref{vdef} that $\hat V^2=0.$ This implies
\begin{equation}
(\hat I+\hat V)^{-1} = \hat I-\hat V \qquad \text{and} \qquad \det(\hat I+\hat V)=1. \label{detid}
\end{equation}
Using the identities~\eqref{detid} we rewrite Eq.~\eqref{detIU} as
\begin{equation}
D(\lambda;n,t) = \det [\hat I+(\hat I-\hat V) \hat W].
\label{DYF}
\end{equation}

The integral operator $\hat W$ can be represented as a product
\begin{equation}
\hat W=\hat A  \hat B,
\end{equation}
where the operators $\hat A$ and $\hat B$ possess the following
integral kernels
\begin{equation}
A(k,q)=\eta(k-q) a(q) \qquad B(q, p)= \eta(q-p) b(q) \label{AB}
\end{equation}
and
\begin{equation}
a(q)=
\left(\begin{array}{cc}
0 & \gamma^* z^{-1}(q)  \\
0 &  \gamma z(q)
\end{array}\right)n_F(q), \qquad
b(q)=
\left(\begin{array}{cc}
0 & 0  \\ z(q) &  z^{-1}(q)
\end{array}\right) n_F(q).
\label{calAB}
\end{equation}
Since $\det(\hat I+\hat A \hat B)=\det(\hat I+\hat B \hat A)$ we rewrite equation \eqref{DYF} as
\begin{equation}
D(\lambda;n,t)=\det [\hat I+ \hat B(\hat I-\hat V) \hat A].
\label{BYA}
\end{equation}
It follows from Eqs.~\eqref{AB} and \eqref{calAB} that
\begin{equation}
\hat B(\hat I-\hat V) \hat A=
\left(\begin{array}{cc}
0 & 0  \\
0 &  \hat K
\end{array}\right), \label{Kfac}
\end{equation}
where
\begin{equation}
\hat K = \hat B_{21} \hat A_{12}+\hat B_{22} \hat A_{22} -
\hat B_{22} \hat V_{21} \hat A_{12}.
\label{Kintermediate}
\end{equation}
The expression on the right hand side of \eqref{Kintermediate} is evaluated with the help of identity
\begin{equation}
\int_{-\pi}^{\pi} \frac{dq}{2\pi} \eta(k-q) \eta(q-p)= \eta(k-p)
\label{iden}
\end{equation}
which gives for the integral kernel of the operator $\hat K$ in Eq.~\eqref{Kintermediate}
\begin{multline}
K(k,p)=[\gamma^* z(k) z^{-1}(p)  +\gamma z^{-1}(k) z(p)] \eta(k-p)\\ +\gamma \gamma^* z^{-1}(k) z^{-1}(p)
 \int_{-\pi}^{\pi} \frac{dq}{2\pi} \eta(k-q) \eta(q-p)  z^2(q).
\label{Kkp}
\end{multline}
The identity
\begin{equation}
\eta(k-q)\eta(q-p)= \eta(k-p)[\eta(k-q)-\eta(p-q)]
\end{equation}
allows us to represent the last term in \eqref{Kkp} as
\begin{equation}
\int_{-\pi}^{\pi} \frac{dq}{2\pi} \eta(k-q) \eta(q-p) z^2(q)=\left\{\frac{1}{2} [z^2(k)+z^2(p)] +f(k)-f(p)\right\}\eta(k-p),
\label{lastterm}
\end{equation}
where
\begin{equation}
f(k) =\frac{1}{2i} \int_{-\pi}^{\pi} \frac{dq}{2\pi}  \frac{z^2(q)-z^2(k)}
{\tan\frac{q-k}{2}}+ \frac{1}{2}\int_{-\pi}^{\pi} \frac{dq}{2\pi} z^2(q).
\label{fdef2}
\end{equation}
Substituting Eq.~\eqref{lastterm} into \eqref{Kkp} we arrive at the expressions \eqref{Kdef0}--\eqref{zdef} for the integral kernel of the operator~\eqref{Kintermediate}. Combining this result with Eqs.~\eqref{BYA} and \eqref{Kfac} we arrive at Eq.~\eqref{main0}.

\section{Conclusion}

Using Wick's theorem we derived a Fredholm determinant representation, Eqs.~\eqref{main0}--\eqref{zdef}, of the time-dependent correlation function~\eqref{corr} of the Jordan-Wigner operator~\eqref{JW}. Generalization to other operators of the Jordan-Wigner type, such as $c_j\mathcal{O}_j,$ and $\varrho_j\mathcal{O}_j,$ is straightforward within this approach. The obtained representation can be used to investigate the charge dynamics in the antiferromagnetic~\cite{ogata_inf, aristov_spinfluct_97, matveev_asymm_zero_bias07, matveev_heisenberg_electrons07}, spin-incoherent~\cite{izergin_impenetrable_fermions, izergin_impenetrable_hubbard, matveev_asymm_zero_bias07, matveev_heisenberg_electrons07, matveev_spin_incoherent_short04, cheianov_spin_decoherent_short, cheianov_spin_decoherent_long, matveev_spin_incoherent_long04, fiete_spin_decoherent, cheianov_momentum_crossover, fiete_SI07}, and spin-polarized~\cite{zvonarev_BoseHubb08, akhanjee_ferrobosons07, matveev_isospin_bosons08, zvonarev_ferrobosons07} 1D gases. It would also be interesting to generalize the calculations given here to the finite-temperature case, in particular, in view of the current interest to the anyon problem~\cite{patu_LenardI08, patu_LenardII08}. Another possible use of the determinant representation is to investigate the applicability range of the low-energy effective field theory (bosonization) description of the functions~\eqref{corr}. The latter description is known~\cite{Aristov_bosonization98} to suffer from Stokes-like phenomena such as the loss of the $\lambda$-periodicity of the large $n,t$ asymptotics of the function~\eqref{corr}. A way to get a $\lambda$-periodic asymptotics could be the use of the technique based on the asymptotic solution of the matrix Riemann-Hilbert problem. This technique, developed in Refs.~\cite{Its_isomono81, Its_Bose90, deift_RH93} proved to be very helpful (see, for example, \cite{cheianov_spin_decoherent_long, kitanine_RH08} and references therein) for the analysis of the asymptotic behavior of the Fredholm determinants with a kernel of the type~\eqref{Kdef}.

\section*{Acknowledgments}

The authors would like to thank V.~E. Korepin and M. Pustilnik for useful comments. This work was supported in part by the Swiss National Science Foundation under MaNEP and division II and by ESF under the INSTANS program.

\end{document}